# BASE DE CONNAISSANCES SYSML POUR LA CONCEPTION DE SYSTEMES COMPLEXES SURS DE FONCTIONNEMENT

# SYSML KNOWLEDGE BASE FOR DESIGNING DEPENDABLE COMPLEX SYSTEM

Romaric GUILLERM et Hamid DEMMOU
CNRS ; LAAS
7 avenue du colonel Roche, F-31077 Toulouse, France
Université de Toulouse ; UPS, INSA, INP, ISAE ; LAAS,
F-31077 Toulouse, France
guillerm@laas.fr, demmou@laas.fr

Nabil SADOU
SUPELEC
IETR
Avenue de la Boulais
F-35510 Cesson Sevigné, France
nabil.sadou@supelec.fr

**Résumé**
Le travail présenté dans cet article s'inscrit dans la proposition d'un cadre de conception des systèmes complexes aussi complet et rigoureux que possible. Le cadre méthodologique utilisé est l'Ingénierie Système, qui est une démarche méthodologique pour maîtriser la conception des systèmes et produits complexes. Les pratiques de cette démarche sont aujourd'hui répertoriées dans des normes, réalisées à l'aide de méthodes et supportées par des outils. Dans notre cas, la norme EIA-632 a été adoptée.
Plus particulièrement, nous nous préoccupons de la sûreté de fonctionnement de ces systèmes complexes. C'est d'ailleurs sur la base de faiblesses des processus actuels traitant de la sûreté que nous avons défini une démarche globale. Cette approche intègre la prise en compte de la sûreté dans les processus d'ingénierie système. Le travail illustré dans cet article vient appuyer et compléter cette démarche globale : il s'agit de la proposition d'un modèle d'information, basé sur le langage SysML, permettant la gestion des exigences, dont les exigences de sûreté.

**Summary**
The work presented in this paper is part of a proposed framework as complete and rigorous as possible for the design of complex systems. The methodological framework used is System Engineering, which is a methodological approach to control the design of complex systems. The practices of this approach are transcribed in standards, realized by methods and supported by tools. In our case, the standard EIA-632 was adopted.
Specifically, to deal with the dependability of these complex systems and to improve the processes dealing with dependability, we have defined a global approach. This approach incorporates the consideration of dependability in system engineering processes. The work presented in this paper supports and complements the overall approach: it is the proposal of an information model based on the SysML language, allowing the requirements management, including safety requirements.

## Introduction

De part leurs tailles grandissantes et leurs fonctionnalités toujours plus sophistiquées, les systèmes actuels sont de plus en plus complexes. Cela influx sur les processus de conception de ces systèmes, qui à leur tour se complexifient. Des propriétés importantes, telles que la sûreté ou la sécurité (Avizienis et coauteurs, 2004), doivent être considérées dès le début de la conception du système. En effet, elles ne peuvent pas être introduites ou seulement mesurées *a posteriori*. Ces propriétés doivent être traitées le plus tôt possible pour limiter leurs impacts sur les délais et les coûts de conception. De plus, elles nécessitent une vue d'ensemble du système pour être traitées, c'est-à-dire une vision globale de tous les sous-systèmes et de leurs interactions.

Les changements apportés aux systèmes au fil des années ont montré les limites des approches et techniques relatives à la sûreté. Ils concernent :
- Des évolutions technologiques rapides,
- Des changements dans la nature des accidents,
- De nouveaux types de dangers,
- Une augmentation de la complexité et de l'hétérogénéité,
- Une relation plus complexe entre l'humain et l'automatisation,
- Un changement de la vision des organismes de régulation et des simples utilisateurs sur la sûreté.

Ces changements représentent un challenge pour tous les acteurs académiques et/ou industriels pour la définition de nouveaux processus, méthodes et outils d'analyse et d'évaluation de la sûreté.

Les échecs d'Ariane 5 et Mars Polar Lander sont des exemples d'accidents système. Dans ces accidents, les composants n'étaient pas défaillants en termes de non satisfaction des exigences pour lesquelles ils ont été conçus. Les composants ont fonctionné exactement comme cela était prévu. Les problèmes proviennent des effets imprévus ou mal compris des comportements des composants sur le système dans son ensemble. Ces erreurs sont des erreurs dans la conception du système plutôt que dans la conception des composants (y compris l'analyse de la sûreté de ces composants), notamment des erreurs dans l'attribution et la traçabilité des fonctions globales du système au niveau des composants individuels.

Les faiblesses des processus d'analyse et d'évaluation de la sûreté actuels peuvent être résumées par les points suivants (liste non exhaustive) (Rasmussen, 1997):
- L'analyse de sûreté comporte un certain degré intrinsèque d'incertitude. Ainsi, il existe un degré de subjectivité dans l'identification des problématiques de sûreté.

- Différents groupes ont besoin de travailler avec différentes vues du système (vue d'ingénieur système et vue d'ingénieur de sûreté par exemple). Généralement c'est un atout, mais peut parfois être une barrière notamment quand les vues sont incohérentes.
- La mauvaise définition des exigences de sûreté et de leur formalisation.
- Absence de traçabilité des exigences de sûreté.
- Les méthodes existantes (traditionnelles) sont insuffisantes vu la complexité des systèmes actuels.
- La description textuelle des modes de défaillance est souvent ambigüe.
- Absence de langage commun entre les différents métiers concernés par le système.

Certains de ces points faibles sont dus à l'absence d'une approche globale d'évaluation de la sûreté. En effet, les propriétés de sûreté sont des propriétés émergentes qui résultent d'interdépendances qui existent dans le système et dans l'interaction avec son environnement. Une solution pour prendre en compte efficacement ces propriétés de sûreté est la proposition d'une approche globale pour l'analyse et la gestion de ces dernières. L'ingénierie système, qui est un cadre méthodologique pour la conception des systèmes complexes, constitue un cadre cohérent et complet pour cette fin. C'est ce cadre que nous considérons dans nos travaux.

Nous avons défini une approche de prise en compte efficace de la sûreté à un niveau global (Guillerm et coauteurs, 2009). Elle intègre les activités d'évaluation de la sûreté dans les activités de développement. L'approche se base essentiellement sur une norme d'ingénierie système, en l'occurrence la norme *EIA-632*, par ailleurs très utilisée. Mais cet article ne se focalise pas sur cette approche générale, qui ne sera donc que sommairement évoquée.

L'objet du papier concerne la proposition d'un modèle d'information basé sur le langage *SysML*, qui permet d'appuyer le déroulement de l'approche mentionnée précédemment. Cette proposition répond à un ensemble de points faibles cités plus haut, dont notamment le besoin d'un langage commun entre les participants d'un projet de conception. Le modèle d'information permet de mieux formaliser les activités de développement et de gestion des exigences de sûreté dans le processus global de gestion des exigences (Juristo et coauteurs, 2002) (Komi-Sirvio et coauteurs, 2003). Il permet une formalisation, une allocation et la gestion de la traçabilité (Gotel et coteur, 1994) (Sahraoui, 2005) des exigences vers le processus de vérification et de validation (V&V).

Ce papier est structuré en quatre grandes sections. La seconde introduit le cadre de conception ainsi que les processus actuels d'évaluation de la sûreté des systèmes complexes. Dans la troisième section, le modèle d'information est proposé, pour une gestion efficace des exigences de sûreté dans le processus de définition et de gestion des exigences. Enfin, la dernière section conclut le propos.

## **Cadre de conception**

### 1    Approche de conception système

Un processus de développement définit les activités nécessaires ainsi que leur ordonnancement pour l'obtention d'un produit final. Deux principales approches ont fait l'objet de nombreux travaux et ont été définies. Le cycle en V (Forsberg et coauteur, 1991) et ses variantes, et l'approche processus.

L'approche processus est basée sur le fait que quelque soit la stratégie utilisée pour le développement d'un système, les activités restent les mêmes. Les processus techniques représentant ces différentes activités d'ingénierie système, sont groupés en deux catégories principales : les processus de définition du système et les processus de vérification et de validation du système. Ils sont définis par de normes d'ingénierie système (IEEE-1220, EIA-632, ISO-15288).

L'approche processus est adoptée dans ce travail car elle est plus flexible que l'approche de développent en V. En effet, la vision processus ne contraint pas la séquence des activités de développement, contrairement aux approches basées sur des cycles (Ménadier, 2002). Cette différence est la motivation pour le choix de l'approche processus, d'autant plus qu'il s'agit de systèmes complexes.

### 2    Ingénierie système

L'ingénierie système fourni un ensemble de concepts facilitant la conception de nouveaux systèmes. Il s'agit d'un processus collaboratif et interdisciplinaire de résolution de problèmes, supportant les connaissances, les méthodes et les techniques résultants des sciences et de l'expérience. L'ingénierie système est en fait un cadre de travail qui aide à définir un nouveau système satisfaisant l'ensemble des besoins des parties prenantes et étant acceptable pour l'environnement, tout en respectant la balance économique globale de la solution pour tous les aspects du problème et pour toutes les phases de développement et de vie du système. L'ingénierie système convient tout particulièrement aux problèmes complexes.

### 3    La norme EIA-632

Un standard notoire, couramment utilisé dans l'industrie ou le secteur militaire, est l'*EIA-632* (Spitzer, 2007). Ce standard couvre le cycle de vie du produit depuis la capture des besoins jusqu'au transfert du système à l'utilisateur. Il fournit une méthodologie d'ingénierie système à travers 13 processus regroupés en 5 catégories (voir la figure 1) : *management technique*, *acquisition et fourniture*, *conception système*, *réalisation du produit* et *évaluation technique*. Un ou plusieurs sous-processus sont définis pour chacun de ces 13 processus et le développeur doit décider lesquels des 33 sous-processus appliquer.

Ce standard fournit un cadre de travail à notre étude, sur lequel s'appuie l'approche de prise en compte de la sûreté au sein de processus généraux d'ingénierie système. Et également, le modèle d'information proposé dans cet article s'inspire de la gestion des exigences définis dans ce standard.

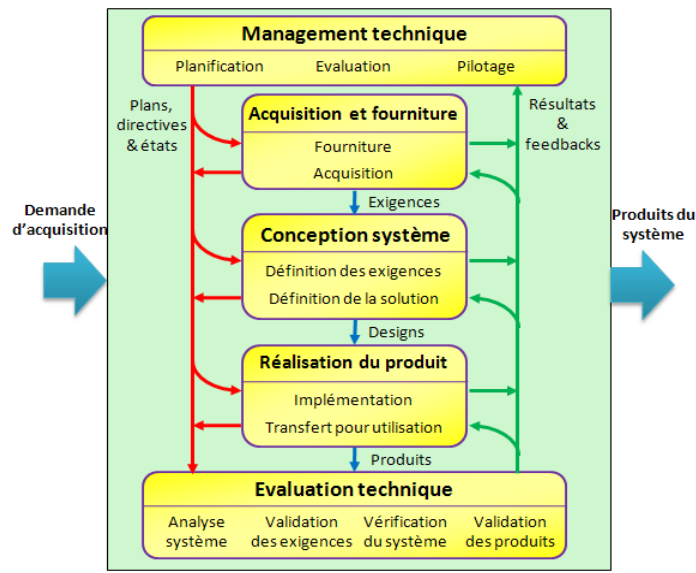

**Figure 1 : Les processus d'ingénierie système de l'*EIA-632***

### 4      Sûreté des systèmes complexes

La sûreté doit être prise en compte dès le début de la conception du système, afin d'éviter des dépassements non-souhaités des coûts et des délais du projet de conception. L'ingénierie système étant un cadre de travail approprié pour la conception de systèmes complexes, nous montrons dans nos travaux comment l'activité d'évaluation de la sûreté peut être explicitement gérée dans ce cadre.

Une approche d'ingénierie système pour la sûreté commence avec l'hypothèse basique que les propriétés de sûreté ne peuvent être traitées entièrement qu'en prenant en compte toutes les variables et tous les aspects (du social au technique) (Kotovsky et coauteur, 1985). Cette base pour l'ingénierie système a été établie à partir du principe qu'un système est plus que la simple somme de ses constituants. En effet, les propriétés de sûreté de fonctionnement des systèmes s'appuient fortement sur les propriétés émergeantes.

La gestion (processus d'évaluation) de la sûreté doit suivre toutes les étapes de l'ingénierie système, depuis la définition des exigences jusqu'à la vérification et la validation du système. Si nous considérons, par exemple, une exigence de sûreté définie et/ou imposée pour le système complet, sa formalisation et son analyse doit permettre d'assurer que les solutions techniques sélectionnées lors de la progression de la conception répondent bien à cette exigence de sûreté au niveau des sous-systèmes et après leur intégration.

Nous avons listé en introduction quelques faiblesses des processus qui influx sur la sûreté de fonctionnement. Présenté dans la partie suivante, le modèle de donnée répond à un certain nombre de ces points, notamment aux besoins d'un langage commun, d'une bonne définition et formalisation des exigences, et d'une traçabilité (Gotel et coauteur, 1995) rigoureuse.

Dans cette première partie, nous avons présenté le cadre Ingénierie Système pour la conception de systèmes complexes. Dans la partie suivante nous allons présenter un modèle d'information basé sur le langage *SysML* pour la modélisation et la gestion des exigences, et en particulier les exigences de sûreté.

## Modèle d'information

### 1      Gestion des exigences

Gérer les exigences dans un projet est une activité fondamentale à son bon déroulement. En effet, un nombre important de documents peut être produit lors de la conception d'un système. Sans la gestion des exigences, il serait difficile de garantir la cohérence et la qualité nécessaire au succès du projet. Effectivement, des études statistiques ont montré que les exigences interviennent pour environ 40% des succès ou des échecs d'un projet.
Ainsi, la gestion des exigences permet de :
- collecter les exigences, et faciliter leur expression,
- détecter les incohérences entres elles,
- les valider,
- gérer les changements d'exigences,
- faire le lien avec le reste du projet et/ou avec le contexte,
- ou encore assurer la traçabilité des exigences.

La gestion des exigences veille aussi à ce que chaque exigence soit correctement déclinée, allouée, suivie, satisfaite, vérifiable, vérifiée et justifiée.

La figure 2 expose une vue de la gestion des exigences de la norme *EIA-632*, sur laquelle s'appuie en partie le modèle d'information proposé plus loin. Les exigences techniques, provenant des exigences des parties prenantes, conduisent aux représentations des solutions logique et physique, qui elles-mêmes sont sources d'exigences techniques dérivées. Puis la solution de conception, répondant aux exigences initiales, est figée par des exigences spécifiées.

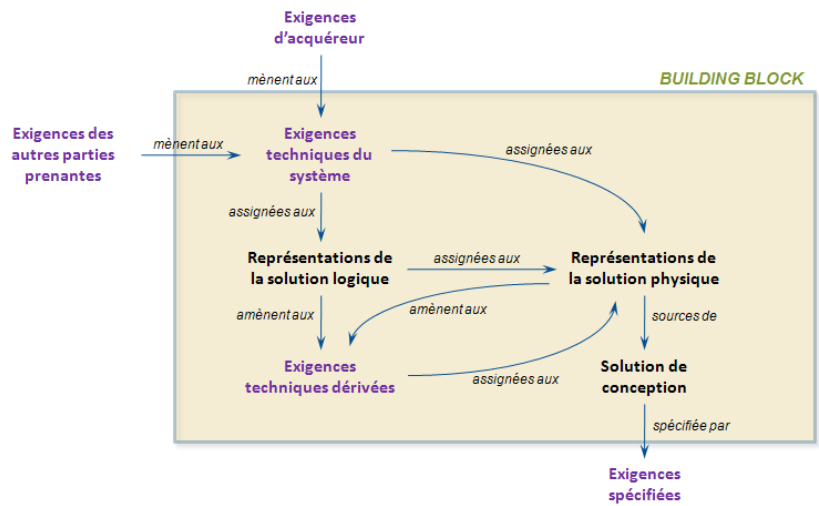

**Figure 2. Gestion des exigences de l'*EIA-632***

## 2    Intérêt d'un modèle d'information

Une manière de rendre efficace la gestion des exigences est de s'appuyer sur un modèle d'information qui serait la base et le cœur de connaissance du projet de conception, sur lequel on va s'appuyer pour :
- guider la conception,
- gérer les modifications/changements d'exigences,
- aider les analyses d'impacts,
- évaluer l'avancement du projet,
- ou plus simplement pour aider à la compréhension du système, ceci sur la base d'un langage commun compréhensible par tous et d'un modèle partagé.

Le modèle d'information proposé dans cet article sera compatible avec les prescriptions de la norme *EIA-632*, tout en y ajoutant des aspects relatifs à la sûreté et la gestion des risques. Ce modèle d'information serait la base des connaissances « système » du projet de conception, permettant le partage des données entre tous les domaines de compétence (mécanique, hydraulique, thermique, électrique, …). Il permet entre-autre le partage des connaissances entre ingénieurs de conception et ingénieurs de sûreté. En effet, la conception de système donne souvent lieu à une accumulation de documentations qui doivent toutes être croisées et mises à jour pour maintenir la cohérence et respecter les spécifications du système. Un modèle d'information est un moyen de regrouper dans un modèle commun à tous les corps de métiers, les spécifications, les contraintes, et les paramètres de l'ensemble du système. Il est donc destiné à modéliser le niveau « système », en exhibant ses interactions avec l'environnement ainsi que les connexions entre les différents sous-systèmes.

En fait, les avantages apportés par un modèle d'information sont nombreux, dont notamment :
- Un partage des spécifications d'un système complexe entre tous les corps de métiers.
- L'identification des risques et la création d'une base d'analyse commune à tous les participants d'un projet.
- Facilite la gestion de projets complexes, l'évolutivité et la maintenabilité des systèmes complexes.
- Documente et capitalise le savoir de tous les corps de métiers dans un projet.

Le modèle d'information doit être vu comme un moyen de mise en commun des connaissances, incluant les 3 volets : exigences, solution de conception et V&V. Il est à considérer comme un véritable niveau d'interconnexion entre les différents métiers, comme représenté dans la figure 3.

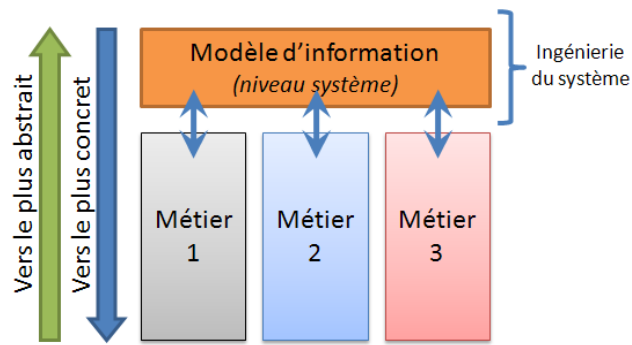

**Figure 3 : Le modèle d'information : un niveau d'interconnexion**

Nous avons choisi le langage *SysML* pour établir ce modèle d'information car la couverture offerte par la diversité de ces diagrammes en fait véritablement le langage pour l'ingénierie système. En effet, *SysML* est un langage de description d'architecture qui convient particulièrement pour les systèmes, de part sa couverture et son abstraction suffisante permettant de décrire aussi bien des solutions logiques (fonctionnelles) que physiques. Basé sur *UML*, il propose en plus la définition d'exigences dans le modèle (à travers le diagramme d'exigences) et couvre également des aspects de traçabilité de ces exigences.

### 3  Langage choisi : SysML

*SysML* (*System Modeling Language*) est à l'ingénierie des systèmes complexes et/ou hétérogènes ce qu'*UML* (Booch et coauteurs, 1998) est à l'informatique. De part la couverture de modélisation offerte par *SysML*, ce langage est un excellent candidat de langage commun.

*SysML* offre tout d'abord la possibilité d'exprimer les exigences à travers le diagramme des exigences. Il définit également plusieurs relations permettant de relier une exigence à d'autres exigences ou à des éléments de modèles. On peut ainsi définir des relations pour la hiérarchie d'exigences, la dérivation d'exigence, le raffinement d'exigence, la satisfaction d'exigence par un élément de modèle ou encore la vérification d'exigence par un cas de test (*TestCase*). Ce langage forme donc une bonne base pour établir notre modèle d'information.

Dans le cadre du processus de définition du système, il est en effet nécessaire d'établir une relation entre les exigences identifiées et les fonctions et/ou composants du système. Ces modèles de traçabilité liant les exigences aux composants du système permettent de réaliser des analyses d'impacts dans le cas d'évolutions du système d'exigences. On est ainsi capable d'évaluer les conséquences de la modification d'une exigence sur la sécurité du système à partir du réseau tissé entre exigences, fonctions et composants.

Pour résumer, les apports de *SysML* à notre démarche d'ingénierie, qui se voudrait donc être une approche d'ingénierie dirigée par les modèles (IDM), sont :

- Langage commun bien défini et compréhensible par tous.
- Permet la modélisation d'une large gamme de systèmes, incluant tant du matériel, que du logiciel, de l'information, des processus, du personnel, ou des équipements
- Traçabilité rigoureuse, ce qui permet de faciliter les analyses d'impacts (exemple : un changement d'exigence)
- Expression soigné des exigences (avec toutes les informations utiles)
- Allocation visible des exigences sur le modèle
- Autre type d'allocation (en plus des exigences) : fonctionnelle et structurelle, ce qui facilite la vérification et la validation
- Bonne définition des interfaces
- Intégration et association des cas de tests directement à la modélisation
- Ajout d'information relative aux risques et propriétés de Sûreté attendues
- Avoir une base de données complète permettant des modifications, des améliorations ou de la réingénierie
- Faciliter la gestion des données (avec un seul langage de modélisation)

### 4  Présentation du modèle d'information

Le modèle d'information (Figure 6) que nous considérons est adapté à la norme EIA-632, notamment en faisant apparaître clairement la distinction entre les différentes classes d'exigences (acquéreur, autres parties prenantes, technique du système et spécifiée).

Pour réalisé ce méta-modèle en *SysML*, nous avons dû étendre le langage. Tout d'abord en définissant de nouveaux stéréotypes pour les exigences (Figure 4), tout en ajoutant nos nouveaux attributs aux exigences, puis en définissant un nouveau type de lien (*specify*) liant les exigences spécifiées aux éléments du modèles.

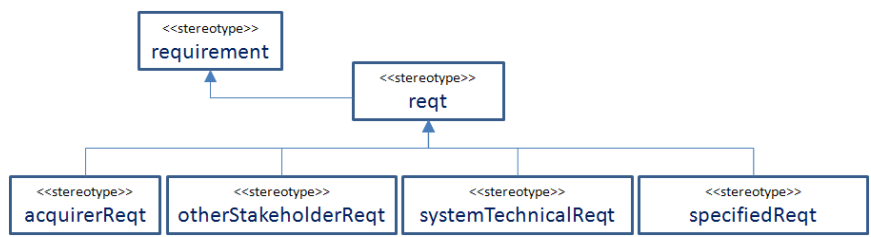

**Figure 4 : Stéréotypes des exigences**

Concernant l'extension du modèle d'exigence (voir figure 5), en plus des champs issus de *SysML* (*ID* et *Description*), nous définissons d'autres attributs essentiels pour une gestion rigoureuse du projet. Ces attributs sont inspirés du profil RPM (Requirement Profile for MeMVatEx) proposé en (Albinet et coauteurs, 2008), des recommandations de l'AFIS (l'Association Française d'Ingénierie Système) et de la *Snow Card Volere (Volere).* Ainsi, les exigences seront spécifiées par leur *catégorie*, leur *source*, leur *cible*, leur niveau de *maturité*, ainsi que leur degré de *criticité*, de *flexibilité* et de *priorité*.

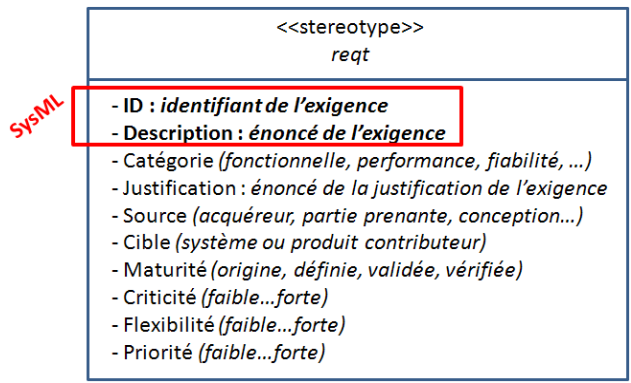

**Figure 5 : Stéréotype étendu d'une exigence**

Dans le modèle d'information, nous avons simplifié le nombre de classes d'exigences en considérant que les « *systemTechnicalReqt* » représentent les exigences techniques du système (évidemment), mais aussi les exigences techniques du système non-allouées à la solution logique et les exigences techniques dérivées provenant de la conception des solutions logiques et physiques.

Les exigences d'acquéreur et des autres parties prenantes sont représentées, sachant que le champ 'source' des exigences doit être en accord avec le stéréotype et désigne plus en détail la partie prenante concernée dans le cas des « *otherStakeholderReqt* ».

Tous les liens de traçabilité demandés par l'EIA-632 sont pris en compte dans ce modèle, et la distinction entre solution logique (partie fonctionnelle) et solution physique (partie composant) apparaît.

Nous mettons en avant dans ce modèle la définition des interfaces, qui sont elles-mêmes des composants et qui lient plusieurs composants entre eux. La notion d'interface est, impérativement, à prendre en compte pour une bonne conception du système. En effet, c'est une des sources de problèmes rencontrés lors de développement.

Le dernier élément important qui est inclus dans ce modèle, faisant ni partie des exigences, ni de la solution de conception, sont les « *TestCase* ». Ces éléments de V&V sont inclus dans le modèle pour être directement liés aux exigences qu'ils vérifient.

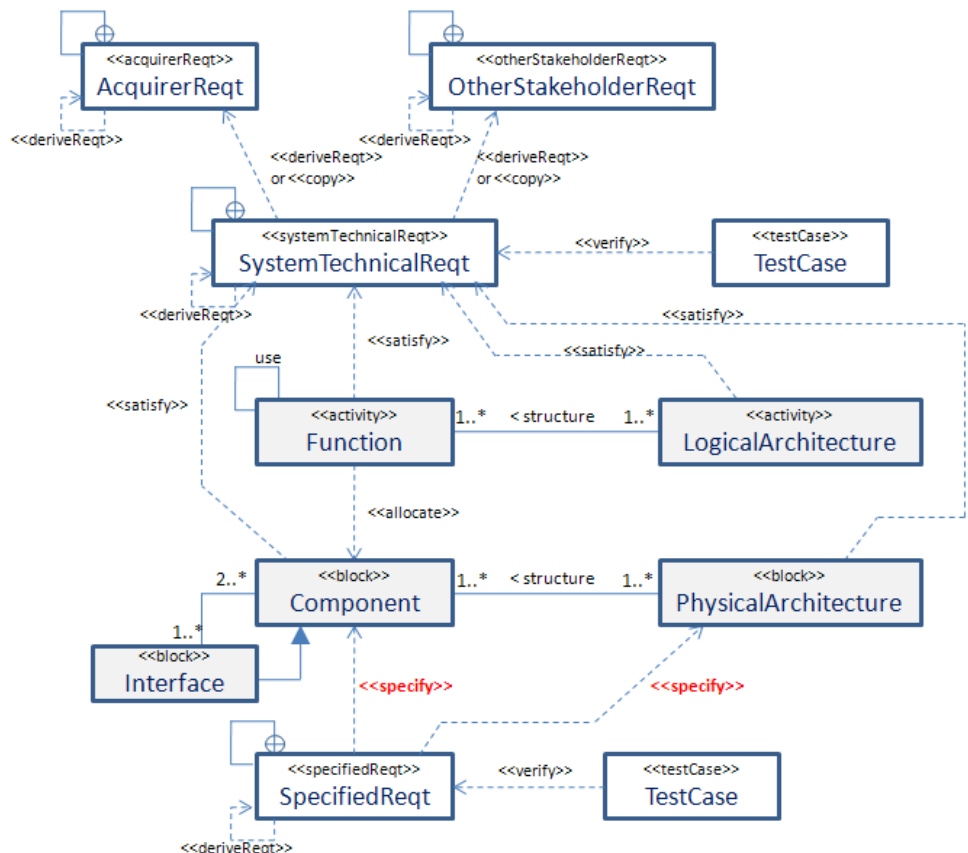

**Figure 6 : Modèle d'information proposé**

# Conclusion

L'ingénierie des exigences est une activité cruciale pour la réussite d'un projet de conception et de développement d'un système complexe. Une gestion efficace des exigences est nécessaire pour mener à bien le projet de développement.

Ainsi, après avoir rappelé le cadre de conception des systèmes complexes, nous avons proposé dans la deuxième partie de cet article un modèle d'information basé sur le langage *SysML*. Pour réalisé ce méta-modèle en *SysML*, nous avons étendu le langage en définissant de nouveaux stéréotypes pour les exigences, tout en ajoutant nos nouveaux attributs aux exigences. Nous avons aussi défini un nouveau type de lien (*specify*) liant les exigences spécifiées aux éléments du modèle. Le modèle proposé permet l'expression de tous les concepts à manipuler, tout en garantissant un cloisonnement entre ces concepts et en permettant la création de lien de traçabilité entre les concepts pour faciliter la compréhension ou les analyses d'impacts.